%% file: paper.tex
\documentclass[conference]{IEEEtran}
\usepackage[pdftex]{graphicx}
\usepackage{color}
\usepackage{multicol}
\usepackage{multirow}
\usepackage{booktabs}
\usepackage{balance}
\usepackage{amsmath,amssymb,amsthm}

\usepackage[normalem]{ulem}
\usepackage{algorithm}
\usepackage{algpseudocode}
\usepackage{subcaption}

\ifCLASSINFOpdf
\else
\fi
\begin{document}
\title{Janus: An Uncertain Cache Architecture to Cope with Side Channel Attacks\vspace{-0.1in}}

\author{\IEEEauthorblockN{Hossein Hosseinzadeh,
Mihailo Isakov,
Mostafa Darabi, 
Ahmad Patooghy and
Michel A. Kinsy}
\IEEEauthorblockA{Adaptive and Secure Computing Systems (ASCS) Laboratory}
\IEEEauthorblockA{Department of Electrical and Computer Engineering, Boston University}\vspace{-0.35in}}


\maketitle

\begin{abstract}
Side channel attacks are a major class of attacks to crypto-systems. Attackers collect and analyze timing behavior, I/O data, or power consumption in 
these systems to undermine their effectiveness in protecting sensitive information. In this work, we propose a new cache architecture, called Janus, to 
enable crypto-systems to introduce randomization and uncertainty in their runtime timing behavior and power utilization profile. In the proposed cache 
architecture, each data block is equipped with an on-off flag to enable/disable the data block. The Janus architecture has two special instructions in its 
instruction set to support the on-off flag. 
Beside the analytical evaluation of the proposed cache architecture, we deploy it in an ARM-7 processor core to study its feasibility and practicality. 
Results show a significant variation in the timing behavior across all the benchmarks. The new secure processor architecture has minimal hardware 
overhead and significant improvement in protecting against power analysis and timing behavior attacks.
\end{abstract}


% \fi
\IEEEpeerreviewmaketitle

\input{intro}
\input{related}

\input{arch}
\input{eval}

\input{concl}

\bibliographystyle{IEEEtran}
 %argument is your BibTeX string definitions and bibliography database(s)
\bibliography{IEEEabrv,paper}
%\begin{thebibliography}{1}
%\end{thebibliography}

\end{document}

%% file: intro.tex
\vspace{-0.05in}
\section{Introduction}
\label{sec:intro}
\vspace{-0.05in}
Computing and embedded systems have penetrated almost every aspect of our daily lives, from mobile phones and artificial pacemakers to 
thermostats and self-driving vehicles. In fact, nowadays, most of the integrated circuits (ICs) in use are found in embedded systems and processing 
sensitive information. The need to improve the security of these systems has never been greater because of the ongoing push to connect them to the 
Internet. To meet some of the security challenges, different crypto-systems have been proposed. However, one of the 
common attacks on crypto-systems, and computing systems in general, is side channel attacks (SCAs) in which external indicators such as power consumption and electromagnetic 
emissions can be used to derive secret and sensitive information. 
Power analysis attacks, fault injection attacks, and timing attacks are among the most successful side channel attacks. With power 
analysis attacks, the power expenditure of a crypto-system is investigated by attackers in order to reveal sensitive information such as cryptographic keys. The most popular power analysis attacks are known as simple power analysis (SPA) and differential power analysis (DPA) attacks \cite{Kocher:1999}. In SPA attacks, the power consumption graphs related to the electrical activities of the IC modules are interpreted 
visually. With DPA techniques, attackers collect and analyze data from various cryptographic functions, and use them to calculate the 
intermediate values of cryptographic computations. 
Since power consumption monitoring is not invasive, the crypto-system may not detect power analysis attacks. To cope with power 
analysis attacks, the system's power consumption can be obfuscated. Randomization of the IC runtime power variations is one such 
technique. By randomizing the consuming power of a crypto-system, attackers find it more difficult to extract secret information. 
Memory operations and the memory hierarchy can be utilized to randomize the power expenditure.
Fault injection attacks are another widely used class of side channel attacks \cite{Ravi:2004}. Fault injection attacks have two main phases.  
In the first phase, the attacker maliciously injects some faults in order to affect the input parameters, processing unit \cite{page2006fault}, storage unit \cite{page2006fault}, or instructions \cite{Yen:2001} of the crypto-system. In the second phase, an analysis is done to gathered information e.g., I/O 
data, timing behaviors to reveal secret keys inside the crypto-system. Fault injection attacks are often based on some well-defined 
analysis vectors \cite{tunstall2011differential} performed on the gathered information during attacks. Randomizing the timing and I/O data of a crypto-system 
significantly improves security of the system, especially, against fault injection attacks \cite{tunstall2011differential}.
In this work, we propose and evaluate a new cache design to cope with power and fault injection attacks. In the proposed \textit{Janus} 
cache architecture, each cacheline has an additional ``on-off flag" (OOF) bit to enable and disable access to the data block. 
%If the the OOF bit is 
%set to zero, the cacheline will be \textsc{off} and a memory request of a words in the data block will result in a miss. If the OOF is one, 
%the associated data block is in the cache and words of the block are made available. 
By introducing instructions to turn on and off cachelines, the runtime power utilization and the timing behavior of the cache structure are efficiently obfuscated. 
%The  rest  of  this  paper  is  organized  as  follows. Section \ref{sec:related} presents the related work. The proposed cache architecture is presented in Section \ref{sec:arch}. Simulation results and comparison with benchmarks are provided in Section \ref{sec:eval}, and conclusions are drawn in Section \ref{sec:concl}. 

%% file: related.tex
\vspace{-0.1in}
\section{Related Work}
\vspace{-0.05in}
\label{sec:related}
%This section introduces previously proposed methods to cope with power analysis and fault injection attacks. 
%Various methods have been proposed to protect against power attacks at different abstraction levels, e.g., transistor and register 
%transfer to system levels. 
In \cite{ambrose2007rijid}, a hardware-software randomized instruction injection scheme (RIJID) was proposed. In RIJID, 
the power utilization is scrambled so that the segments of the encryption code cannot be identified. The scheme has shown some efficacy 
against both SPA and DPA attacks that use system power profile to extract encryption code. 
Ambrose \emph{et al.} in \cite{ambrose2013randomised} proposed the use of parallel capability in multi-modulo residue number systems (RNS) architectures 
to scramble sensitive data. By using RNS architectures, the operations can be divided into parallel sections, and thus, the power 
consumption and complexity are reduced. Yang \emph{et al.} \cite{Yang:2005} introduced a scheme known as random dynamic voltage and 
frequency scaling (RDVFS) to decrease the correlation between the system power consumption and input data by changing the 
frequency and voltage randomly. However, RDVFS method cannot  defeat  SPA/DPA  power  attacks \cite{Yang:2005}. In \cite{avirneni2014countering}, authors 
developed a policy using dynamic voltage and frequency scaling (DVFS) to overcome the limitations of RDVFS by breaking correlations 
between voltage and frequency of (V, f) pairs. In \cite{masoumi2015novel}, the advanced encryption standard (AES) algorithm is implemented using 
techniques resistant to first order differential electromagnetic and power analyses. With this approach, the Galois Field of the AES is 
randomized and no additional operation is added to the algorithm. Consequently, the working frequency remains the same and the used 
algorithm is compatible to the published standard.
Fault attack tolerant methods generally fall under one of two categories: fault avoidance and fault protection. In both cases, extra hardware is often required 
to (a) check and prevent fault injection or (b) rollback the crypto-system to recover from the fault. 
%%In the former case, proposed methods 
%%\cite{tunstall2011differential} try to prevent the fault injection phase of the attack. Extra hardware are often utilized to check and prevent fault injection. 
%%Examples of this are active and passive shielding methods \cite{bar2006sorcerer}, usage of fault detecting sensors are examples of this approach. 
%%These approaches often lead to high implementation cost and performance overhead. Furthermore, they still fail in to protect crypto-systems against new methods of fault injection attacks. 
Most of the proposed approaches dealt with power attacks while ignoring fault attacks or vice versa. In this work, we try to jointly address both fault injection 
and power attacks in crypto-systems. 
%The proposed method is described in following sections. 

%% file: arch.tex
\vspace{-0.05in}
\section{Janus Cache Architecture}
\label{sec:arch}
\vspace{-0.05in}

In both general-purpose and embedded computing, the overall system's performance and power usage is highly dependent 
on the cache's performance. When the processor needs some data, it goes to the cache. If the data is in the cache, there is a hit. 
Otherwise, the processor has to wait for main memory to supply the data. Since access time for the main memory is orders of 
magnitude greater than the cache access time, cache hits and cache misses have very different access times and power profiles. 
From the power consumption view, a cache hit consumes very little energy since no external lines of data are moved through the 
memory subsystem hierarchy and no main bus address or data activities are involved. Therefore, the hit rate of the cache system 
plays a pivotal role in the power consumption and timing behavior of a crypto-system. The key insight is that by changing the miss 
and hit rates, one can alter the power consumption and timing behavior of the system leading to a more robust crypto-system.

The proposed cache architecture operates under fully associative policy for substituting the data words.  More specifically, new data 
words can be stored in any free locations of the cache, and if the cache is full, data eviction and new data words placement use the 
Least Recently Used (LRU) policy. In the \textit{Janus} cache design, for each block of data there is one flag bit called ``on-off flag" 
(\textsc{oof}). The \textsc{oof} is used to enable or disable access to a particular cacheline even when the valid bit of the line is one 
and there is a match on the tags. By introducing a small set of instructions for turning on and off the \textsc{oof} bits, we are effectively 
able to (a) obfuscate the power utilization of system in a controlled manner and (b) minimize the hardware modifications needed to 
support the new security feature. 
%%Figure \ref{fig1} shows the structure of \textit{Janus} architecture. 
All the fields in the conventional cache 
structure and their functionalities remain the same. For simplicity, we did not show the cache coherence bits field. The \textsc{oof} bit 
check happens after the valid bit check, therefore, in the \textit{Janus} architecture, there is one single gate delay in the cache structure. 
%\begin{figure}
%  \includegraphics[width=1.5in]{figs/fig1.pdf}
%  \centering
%  \vspace{-0.05in}
%\caption{The proposed cache design for a direct mapped structure.} 
%\label{fig1}
%\vspace{-0.2in}
%\end{figure}
To control the state of each \textsc{oof} of the cache structure, we introduce two \textsc{on}-\textsc{off} instructions: ``cache-block-on-$i$" and ``cache-block-off-$i$", 
for controlling the $i$-th cacheline. 
%For example, by executing the ``cache-block-off-7" renders the 7th line unusable even if it resides in the 
%cache until the ``cache-block-on-7" is executed to revert the effect. 
By exploiting these two instructions, the amount of effective hit and miss 
rates of the cache is controlled beyond the normal miss and hit rates of the executing program. This approach gives users a program level access 
for controlling the desired amount of obfuscation. The random injection of this instruction pair into the based code creates a runtime power profile 
and timing behavior for the crypto-system that are more resilient to power analysis and fault injection attacks.

%\vspace{-0.05in}
\subsection{Runtime ON-OFF Algorithm}
\vspace{-0.05in}
%Here we introduce Figure \ref{fig2} to help illustrate the \textit{Janus} architecture runtime ``on-off" policy. Each memory request now has three 
%possible outcomes (1) program miss, (2) program hit and \textsc{oof} \textit{off}, and finally (3) program hit and \textsc{oof} \textit{on}. Case (2) 
%is called an intentional miss (IM). 
%When the addressed block has not been previously brought in the cache or has been evicted, a cache miss occurs, e.g., accessing block 1, 
%2, 3 in time slots 1, 2, 5. However, if the block is found in the cache, there is a cache hit, e.g., accessing block 1, 2, 4 in time slots 3, 4, 8. 
%Since cachelines may be disabled by the user, certain cache misses are intentional (IM), shown in Figure \ref{fig2}. When in time slot 7 the block 
%2 is disabled an IM ensues. In the case of an IM, the crypto-system follows the same data fetching process (either from lower caches or main 
%memory) as in the case of a genuine cache miss. To make sure that an IM and an actual miss have the same power and latency profiles, the 
%fetched block is placed on top of the old. 
%
%\begin{figure}
%  \includegraphics[width=3.0in]{figs/fig2.pdf}
%  \centering
%  \vspace{-0.1in}
% \caption{An example of address sequence under the Janus caching scheme. }
% \label{fig2}
% \vspace{-0.2in}
%\end{figure}

Each memory request now has three 
possible outcomes: (1) program miss, (2) program hit and \textsc{oof} \textit{off}, and finally (3) program hit and \textsc{oof} \textit{on}. 
When the addressed block has not been previously brought in the cache or has been evicted, a cache miss occurs. However, if the block is 
found in the cache, there is a cache hit. Since cachelines can be disabled through the security policy, certain cache misses are intentional (IM) 
- outcome (2). In the case of an IM, the crypto-system follows the same data fetching process (either from lower caches or main 
memory) as in the case of a genuine cache miss. To make sure that an IM and an actual miss have the same power and latency profiles, the 
fetched block is placed on top of the old. 
Let us assume that the cache has $n$ data blocks, from 0 to $n-1$, and the considered code to be run consists of $m$ time slots, from 0 to 
$m-1$. Turning off each data block increases the power utilization of the crypto-system. This execution time overhead is based on the 
amount of IMs encountered during program execution. This increase in power can be modeled as a random variable, more 
specifically as a Poisson random variable, since it depends on the number of \textsc{on}-\textsc{off} instruction 
pairs executed at runtime. 
Let $P_i$ denote the increase in power in the crypto-system when the $i$-th cache data block is turned off. Thus, for $P_i$, we have: 
\vspace{-0.05in}
\begin{equation}
P_i=A_i\times C_0,
 \label{eq1}
 \vspace{-0.05in}
\end{equation}
$A_i$ is the number of active requests on the $i$-th data block during execution and $C_0$ is a constant value (the power consumed by the 
crypto-system to bring in data from the RAM - Random Access Memory - instead of the cache). 
With the Janus caching scheme, the execution of \textsc{on}-\textsc{off} instruction pairs and their effects on the 
cache miss rate add uncertainty to the power consumption of the crypto-system, and obfuscate the actual program execution power usage 
profile. As a result, the crypto-system is protected against the power analysis attacks. 
The runtime power utilization uncertainty or the added noise is a \emph{random process}. In practical systems, the power consumption is 
capped (i.e., the second moment of the noise is limited), therefore, the highest uncertainty (i.e., entropy) in the power consumption can 
be realized with a Gaussian noise model \cite{proakisdigital}. For this reason, the Janus caching scheme creates a Gaussian noise in the power 
consumption through the random variable $P_i$ and uses it to insert the appropriate number of \textsc{on}-\textsc{off} instruction pairs in the code. 
Let $n(t)$ denote the amount of Gaussian noise at the time slot $t$  which can be modeled by a Gaussian \emph{random variable}. 
Because of practical limitations, a pure Gaussian random variable cannot be generated, thus, a pseudo Gaussian random variable 
at the time slot $t$ is used. For producing the $n(t)$, at first, we choose two numbers $U_1$ and $U_2$ in the range of $[0,1]$ 
arbitrarily. $n(t)$ can be produced using the following equations \cite{proakisdigital}:
\begin{IEEEeqnarray}{lll}
V_1=2U_1-1 ~~~~~~ and ~~~~~~~V_2=2U_2-1, \nonumber\\
S={V_1}^2+{V_2}^2, \,\,\,\, (\text{Such that}\,\, S\leq1), \nonumber\\
n(t)=\sqrt{\frac{-2\log(S)}{S}}V_1
 \label{eq2}
\end{IEEEeqnarray}
If $S>1$, we select another $U_1$ and $U_2$ until $S\leq1$ holds. This is the amount of Gaussian noise at the time slot $t$ that should 
be added to the power consumption of the system. Algorithm~\ref{algo1} presents the procedure for achieving the value of $n(t)$ in Equation~\ref{eq2} 
to be added to the random variables $P_i$ in Equation~\ref{eq1}. The algorithm derives in $n(t)$ by exploring the \textsc{on} and \textsc{off} 
states of the cache data blocks in the time slot $t$.
 \vspace{-0.05in}
\begin{algorithm}
\caption{Janus \textsc{On}-\textsc{Off} Policy in the Time Slot $t$}
\begin{algorithmic}[1]                                       
\State  Compute all the $P_i$ values via Equation~\ref{eq1}.
\State  Compute the power addition or minus for all the states of data blocks.
\State Compute $n(t)$ value via Equation~\ref{eq2}.
\State  Among the different states of data blocks for being turned on and off, choose the state whose result is the closest to the amount of $n(t)$ computed in the previous step. 
\end{algorithmic}
 \label{algo1}
\end{algorithm}
It is worth noting that, although we illustrate the Janus architecture with a single level cache for presentation simplicity, it works in multi-level cache 
systems as well. 

%% file: eval.tex
\vspace{-0.07in}
\section{Evaluations}
\label{sec:eval}
\vspace{-0.05in}

\subsection{Analytical Evaluation}
 \vspace{-0.05in}
The analytical assessment of the Janus cache architecture focuses on (a) the number of available data blocks to turn \textsc{on} and 
\textsc{off} in the cache at any given time and (b) the error probability of guessing the consumed power. 
Let us assume that there are $N$ data blocks available to turn \textsc{on} and \textsc{off}, and turning off each of them results in 
some $P$ power increase. During each time slot, there are $N+1$ possible data block states and $\frac{2NP}{N+1}$ power difference 
between them. Based on \cite{Tilborg:2005} noise quantization results, the variance of distance between the Pseudo-Gaussian noise produced by 
turning \textsc{on} and \textsc{off} the data blocks and the Gaussian noise, i.e., $\sigma_{\texttt{distance}}^2$, can be calculated as
\begin{equation}
 \vspace{-0.05in}
\sigma_{distance}^2=\frac{NP^2}{3(N+1)^2}.
\label{eq3}
% \vspace{-0.05in}
\end{equation}
By increasing the number of available data blocks in the cache structure to be turned \textsc{on} and \textsc{off}, one can decrease the 
distance between the two noises. 
Another important metric for evaluating the Janus architecture's performance is its error probability in estimating the crypto-system's power 
usage to the \textsc{on} and \textsc{off} decisions. Since a Gaussian noise model is used, one can model the error probability of the power 
estimate as the error rate in an \textit{additive white Gaussian noise} (AWGN) channel using \textit{binary phase-shift keying} (BPSK) 
modulation. Using the same approach as in \cite{Tilborg:2005} for AWGN channel, the error probability of the estimated power, i.e., $Pr_{error}$, can 
be written as
\begin{equation}
 \vspace{-0.05in}
Pr_{error}=\frac{1}{2}erfc(\sqrt P)
\label{eq4}
% \vspace{-0.05in}
\end{equation}
Where $erfc$ is the error function and equals to $\frac{2}{sqrt{\pi}}\int_{x}^\infty e^{-x^2}dx$.

\subsection{Simulation Results}
\begin{figure*}
    \centering
    \begin{subfigure}[b]{0.45\textwidth}
        \includegraphics[width=2.8in]{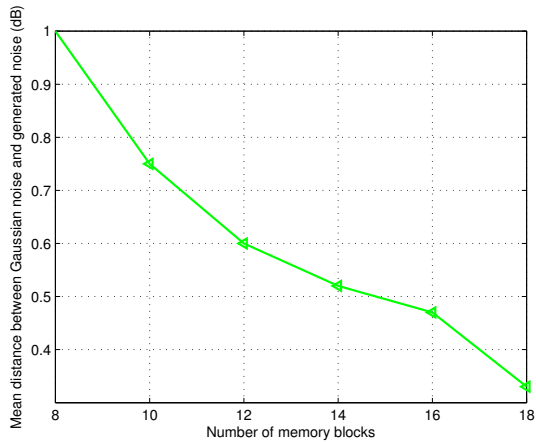}
        \caption{Mean distance between the Gaussian noise and the generated noise versus the number of data blocks in the cache structure.}
        \label{fig3}
    \end{subfigure}
    \hspace{.15in}
    \begin{subfigure}[b]{0.45\textwidth}
        \includegraphics[width=2.8in]{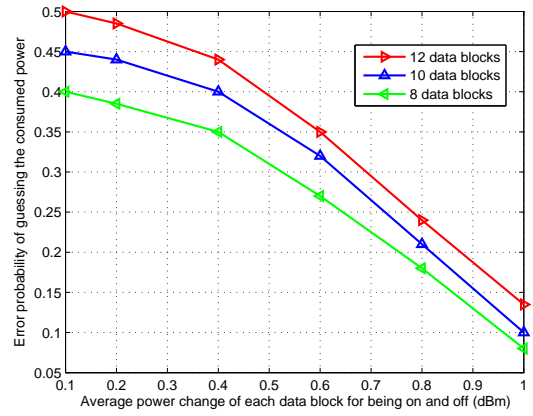}
        \caption{Error probability of the power estimate versus the average power change due to data blocks being turned \textsc{on} or \textsc{off}.}
        \label{fig4}
    \end{subfigure}
    \vspace{-0.05in}
    \caption{Results of the mean distance and error probability.}
    \label{fig34}
    \vspace{-0.05in}
\end{figure*}

\begin{figure*}
    \centering
    \begin{subfigure}[b]{0.45\textwidth}
        \includegraphics[width=\textwidth]{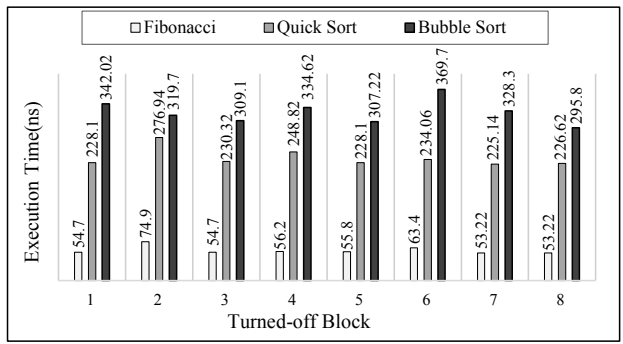}
        \caption{Under random \textsc{on}-\textsc{off} scheme.}
        \label{fig5}
    \end{subfigure}
    \hspace{.1in}
    \begin{subfigure}[b]{0.425\textwidth}
        \includegraphics[width=\textwidth]{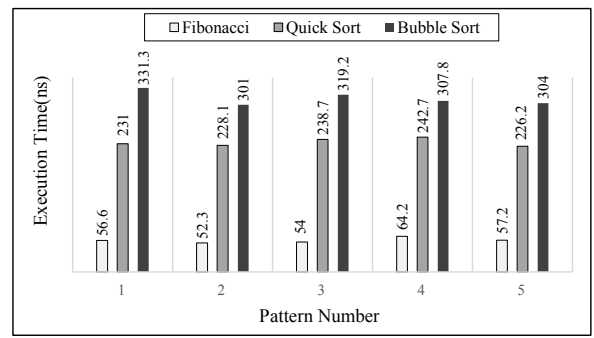}
        \caption{Under the five predetermined \textsc{on}-\textsc{off} patterns.}
        \label{fig7}
    \end{subfigure}
    \vspace{-0.05in}
    \caption{Runtimes for the benchmarks under different \textsc{on}-\textsc{off} schemes.}
    \label{fig57}
    \vspace{-0.25in}
\end{figure*}

 \vspace{-0.05in}
%The Janus cache architecture is implemented and under various modalities, the data blocks are turned \textsc{on} and \textsc{off} using 
Data blocks are turned on and off in the Janus cache architecture using Algorithm \ref{algo1}. We compare the measured consumed power of the 
system to the theoretical Gaussian noise model to show that the proposed 
architecture effectively randomizes the power consumption of the crypto-system. 
Figure~\ref{fig3} shows the mean distance between the Gaussian noise and the produced noise under Janus' \textsc{on} and \textsc{off} of 
data block policy. By increasing the number of the data blocks in the cache structure, the average distance between the produced noise 
and the Gaussian noise is reduced. 
%When the number of data blocks available for the \textsc{on} and \textsc{off} policy execution is increased, 
%the distribution of the produced noise gets closer to the Gaussian noise and makes power attacks on the crypto-system more difficult. 
Figure \ref{fig4} shows the error probability of the estimated power usage as a function of the average power change introduced by data blocks 
being turned \textsc{on} or \textsc{off}. By increasing the number of data blocks and the average change power of each data block, the error 
probability of power consumption estimate decreases.
%\begin{figure}
%  \includegraphics[width=2.0in]{figs/fig3.pdf}
%  \centering
%  \vspace{-0.1in}
% \caption{Mean distance between the Gaussian noise and the generated noise versus the number of data blocks in the cache structure.}
% \label{fig3}
% \vspace{-0.05in}
%\end{figure}
%\begin{figure}
%  \includegraphics[width=2.0in]{figs/fig4.pdf}
%  \centering
%  \vspace{-0.1in}
%\caption{Error probability of the power consumption estimate versus the average power change due to data blocks being turned \textsc{on} or \textsc{off}.}
% \label{fig4}
% \vspace{-0.05in}
%\end{figure}
To fully evaluate the concrete and practical implications of the Janus cache architecture on the timing behavior of a crypto-system, we 
deploy the Janus architecture in a gate-level synthesized version of the ARM-7 processor that is simulated using the XILINX ISIM simulator. The timing behavior of the synthesized ARM-7 on three 
benchmarks (1) Fibonacci sequence generator, (2) quick sort, and (3) bubble sort is extracted and analyzed.
The timing behavior results are reported in Figure \ref{fig5}. Depending on the temporal locality of the program, turning off a single 
data block in the cache can have a significant effect on the runtime behavior of the system. 
%For instance, in bubble sort benchmark, turning 
%off the 6th data block and turning off the 8th data block have the most and least effects on the system runtime behavior. 
%For the quick sort benchmark, turning off the 2nd data block and turning off the 7th data block have the most and least effects on the runtime 
%behavior. 
A powerful resulting insight from this analysis is the fact that even if an attacker identifies the data block with the most effect on the system 
runtime behavior under one program code, this information may not be useful or effective in attacking the same crypto-system running another 
program. Therefore moving target security features are also present in the Janus design as a byproduct.\\

%\begin{figure}
%  \includegraphics[width=3.0in]{figs/fig5.pdf}
%  \centering
%\vspace{-0.05in}
%\caption {Runtime of bubble sort benchmark, quick sort benchmark, and Fibonacci benchmark for turning off a data block randomly (in nanoseconds time intervals).}
% \label{fig5}
% \vspace{-0.1in}
%\end{figure}
%
%%\begin{figure}
%%  \includegraphics[width=3.0in]{figs/fig6.pdf}
%%  \centering
%%  \vspace{-0.05in}
%%\caption {The method for turning on and off the data blocks in cache.}
%% \label{fig6}
%% \vspace{-0.05in}
%%\end{figure}
%
%\begin{figure}
%  \includegraphics[width=3.0in]{figs/fig7.pdf}
%  \centering
%  \vspace{-0.05in}
%\caption {Runtime of bubble sort benchmark, quick sort benchmark, and Fibonacci benchmark for turning off a data block randomly according to method in Figure \ref{fig6}}
%  \label{fig7}
%  \vspace{-0.05in}
%\end{figure}

\begin{table}[h!]
\vspace{-0.1in}
  \centering
  \caption{The mean and variance of the runtimes presented in Figure~\ref{fig5}.}
    \vspace{-0.05in}
  \label{tab:table1}
   \begin{tabular}{ccc}
    \toprule
     & Mean execution & Execution time \\
     &    time (ns)      &    variance\\
    \midrule
    Bubble sort & 325.8 & 547.951\\
    Quick sort & 237.26 & 313.49\\
    Fibonacci & 58.26 & 55.67\\
    \bottomrule
  \end{tabular}
  \vspace{-0.05in}
\end{table}

\begin{table}[h!]
\vspace{-0.05in}
  \centering
  \caption{The mean and variance of the runtimes presented in Figure~\ref{fig7}.}
    \vspace{-0.05in}
  \label{tab:table2}
   \begin{tabular}{cccc}
    \toprule
     & Mean execution & Execution time & Normal execution  \\
     &     time (ns)     &    variance    &     time (ns)     \\
    \midrule
    Bubble sort & 311.86 & 122.11 & 290.220\\
    Quick sort & 238.12 & 88.65 & 223.660\\
    Fibonacci & 58.26 & 55.67 & 51.740\\
    \bottomrule
  \end{tabular}
  \vspace{-0.1in}
\end{table}

To investigate the effect of turning on and off the data blocks on the program execution profile, we create 5 different patterns, each pattern has 5 time 
intervals (the first four are 60 nanoseconds long and a last interval runs to the end of the program). For each pattern, at each interval, different cache 
block sets are turned on and off. Figure~\ref{fig7} shows the results of these experiments. The mean and variance of the runtimes for the benchmarks 
in Figure~\ref{fig5} are summarized in Table~\ref{tab:table1}. Table~\ref{tab:table2} presents the mean and variance of runtimes for benchmarks under 
the different \textsc{on}-\textsc{off} patterns. 
The results show that even under this simple time slicing approach, the Janus architecture scrambles the mean and variance of the program runtime 
enough to provide strong protection against fault injection attacks. 

%In this table normal execution time which is the time needed to run a benchmark without any cache block on/off is also reported to show impacts of the proposed cache architecture.

%% file: concl.tex
\vspace{-0.1in}
\section{Conclusion}
\label{sec:concl}
\vspace{-0.05in}
In this work, we propose a new caching architecture, called Janus, to enable the randomization of the power consumption in crypto-systems. By 
obfuscating the runtime power profile the Janus architecture is able to effectively protect these systems against power analysis and timing behavior 
attacks. The Janus cache architecture is deployed in a synthesized ARM-7 processor core running three different benchmarks to evaluate (a) the 
feasibility of the architecture, and (b) its efficacy against the mentioned attacks. 
%%Results confirmed that the power behavior of the designed 
%%crypto-system is very close to the Gaussian noise model with strong protection against power analysis attacks. 
 \vspace{-0.1in}